# Institutionalizing Best Practices in Research Computing: A Framework and Case Study for Improving User Onboarding

AYUSH CHATURVEDI, ROB POKORNEY, ELYN FRITZ-WATERS, CHARLOTTE ROUSE, GARY BAX, DARYL SPENCER, and CRAIG POHL, Washington University in St. Louis, USA

Research computing centers around the world struggle with onboarding new users. Subject matter experts, researchers, and principal investigators are often overwhelmed by the complex infrastructure and software offerings designed to support diverse research domains at large academic and national institutions. As a result, users frequently fall into confusion and complexity to access these resources, despite the availability of documentation, tutorials, interactive trainings and other similar resources. Through this work, we present a framework designed to improve new-user onboarding experience. We also present an empirical validation through its application within the Research Infrastructure Services at Washington University in St. Louis.

## 1 Introduction

Research computing centers (RCCs) or high-performance computing (HPC) centers are critical to the scientific community. Millions of researchers and students worldwide rely on critical computing infrastructure to advance science. Thus, it is evident that accessibility and rapid onboarding to RCC services mandate the least amount of time, effort, and complexity. However, this seldom happens due to computer science and engineering principles being rooted in the RCC staff and foundation; after all, they provide computing services. As a result, end users: researchers, principal investigators (PIs), and students find themselves lost in a set of complex documentation, video libraries, and deviations in RCC infrastructure as compared to their local workstation. A similar case was in our center, where users often complained about a poor onboarding experience. Hence, we decided to overhaul all aspects and entry points of user onboarding journey. We carried out our efforts for over a year, in phases. With some success, we share our journey and a novel framework dedicated to improving user experience (UX) in HPC and RCC centers in this work. Below, we outline background of our center and its user followed by problems in our onboarding services.

Authors' Contact Information: Ayush Chaturvedi, ayush@wustl.edu; Rob Pokorney, rpokorne@wustl.edu; Elyn Fritz-Waters, elyn@wustl.edu; Charlotte Rouse, rouse@wustl.edu; Gary Bax, bax@wustl.edu; Daryl Spencer, daryls@wustl.edu; Craig Pohl, cpohl@wustl.edu, Washington University in St. Louis, St. Louis, Missouri, USA.

### 1.1 WashU-RIS Background

Washington University in St. Louis (WashU) maintains a prolific research ecosystem with landmark contributions to genomic sequencing [16], metabolic microbiome function [33], Alzheimer's disease progression [4], and Martian habitability [14]. Sustaining an annual research portfolio exceeding one billion dollar necessitates sophisticated computational expertise; however, these efforts were historically distributed and domain-focused. As a result, several internal collaborations formed with evolution of institutional computing capabilities over thirty years.

In 2017, WashU leadership formed the WashU Information Technology (IT) research computing department and WashU IT Research Infrastructure Service (RIS) group. Originating from the Human Genome Project [16] and part of the McDonnell Genome Institute, WashU IT RIS brought a strong background in genomics workflows, high-speed data transfers, high-throughput computing, container technologies, multi-petabyte parallel filesystems, and security to support both basic science and clinical research efforts.

### 1.2 WashU-Researcher Background

WashU is primarily composed of students, researchers, faculty and PIs working at two physical campuses separated by a large urban park. On one side are the medical schools and the large urban hospital, on the other are undergraduate and graduate schools. Due to such diverse user community, WashU lacked common central services until the formation of WashU IT in 2014, which consolidated IT organizations and services across both campuses. Despite central IT services, until RIS began operations, departmental IT teams and researchers were themselves responsible for technological solutions for their programs. This resulted in hundreds of different technology solutions, operational standards, and funding mechanisms.

### 1.3 Motivation

In 2018, RIS released its first 'Storage Service' accessible from a network share or an HPC cluster followed by release of integrated HPC services in 2020. Onboarding users required researchers and students to submit a ticket for 'Storage' or 'Compute' services. Provisioning access and providing a list of instructions and references to our documentation was a standard onboarding procedure. Provisioning was catered through internal ITSM portal supported by Jenkins. For technology support, we leverage JIRA for service desk, request management and project management.

Users were expected to understand our architecture and technology stack, be willing to learn by consulting our documentation, and understand the gap between our services and determine how to use them effectively. However, as evident from Table 3, user who request computing infrastructure services at RIS lack a formal computer science or engineering background (only ≈ 8%sponsors and ≈ 6% departments). As a result, users reported a lack of personalized support, complex navigation, and insufficient guidance, resulting in a steep learning curve and user dissatisfaction. Addressing these shortcomings was imperative to improve user retention and satisfaction and prompted us to revamp our onboarding processes to ensure a more intuitive, efficient, and supportive user experience. Consequently, we ran a project for one complete year to improve onboarding experience which we share as framework for the broader community.

## 2 Prior Works

Computing centers around the world have made many efforts to enhance the user onboarding experience. Julich supercomputing center [37] created dedicated courses similar to PNNL [10], while University of Colorado Boulder [18, 26] improved documentation, website and tutorials in an attempt to improve user experience. They also highlight the role of community engagement and collaborations with other local computing consortium. Although this is a great way to educate users from non-computer-science backgrounds, it may not be as useful much for users with a computer-science background. Other works such as [1, 35] attempts to provide similar frameworks as ours, however, they only cover either domain specific areas to onboarding or deviates from onboarding in their primary focus to beyond. Moreover, both lack testimonial evidence supported by data.

## 3 Framework for Improving New User Experience

With increasing complexity in infrastructure and services around it, in RCCs, the entry point for researchers often becomes opaque and confusing. This results in unwarranted speed bumps for 'research computing users', and in turn, delays essential scientific discovery.

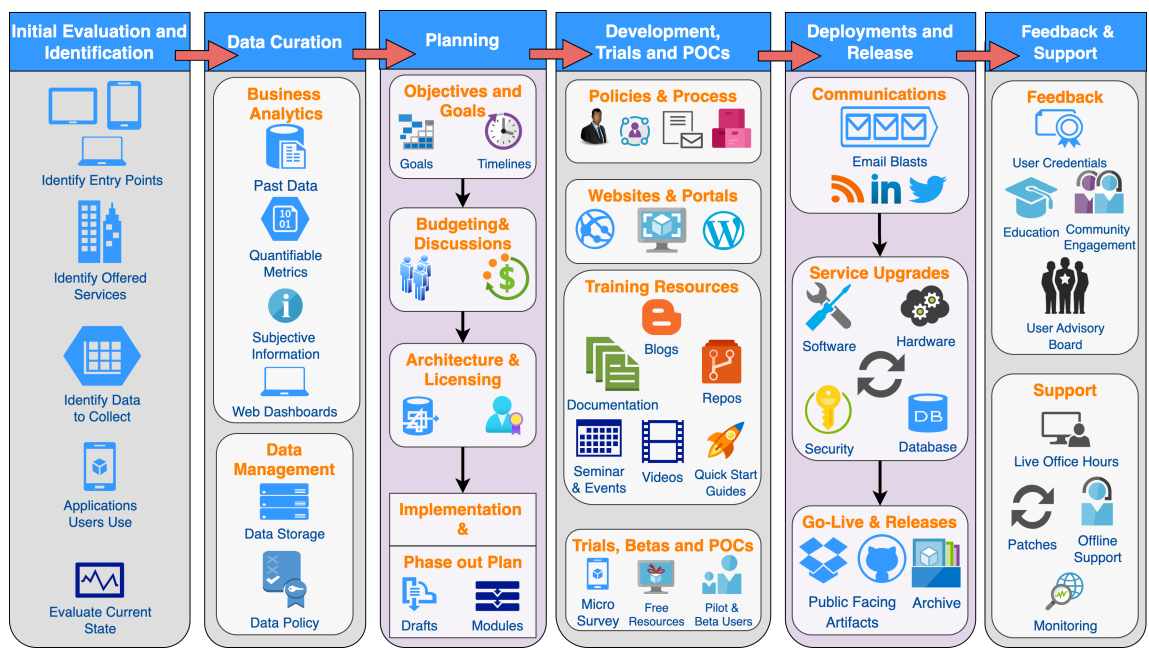

Fig. 1. Overview of Framework for Improving User Experience at a Research Computing Center.

Therefore, we introduce a comprehensive framework designed to replace ad-hoc methods and philosophies with a streamlined, robust, and articulated set of processes to help improve user onboarding experience. Figure 1 shows a detailed single-point bird's eye view of our framework that consists of six phases.

By institutionalizing this blue print and practices outlined in this work, computing centers can drastically reduce the user learning curve and administrative overhead for staff. Our pioneering work converges multiple aspects at a computing center to a single goal aimed to champion overall user experience. Principles and method we share can minimize 'time-to-science', ensuring that researchers can transition from registration to computation with efficiency and confidence. Below, we elaborately discuss each part in this framework.

## 3.1 Initial Evaluation and Identification

Initial evaluation for current state of affair builds a robust foundation for planning and evaluation. Before any onboarding interventions can be designed, a research computing center (RCC) must establish a baseline understanding of current user ecosystem with objective and subjective evaluation. Additionally, drawing from information technology service management (ITSM) [9, 22] and user experience (UX) [8, 27, 32] postulates, RCC staff should try to quantify and diagnose friction points and align service delivery with actual 'research user' needs. The process and methods for identification and evaluation can be outlined in three critical categories as addressed below.

*3.1.1 Identify Entry Points.* An RCC ecosystem always has multiple entry points for users to interact with the offered services. Mapping the user journey requires identifying every touch point where a current or potential user may interact or request services for the first time. These often include RCC's website, ticketing portal, and other gateways and service portals, such as Open-OnDemand, ColdFront, and XDMoD, etc. Other indirect entry points can originate from RCC collaborations and science gateways. For e.g., an NSF NAIRR Pilot user may choose to use Jetstream2 cloud infrastructure services and the 'Exoshpere' [15] science gateway. Therefore, identification of all direct and indirect entry points *early on* is critical. Formalizing these entry points ensures consistent information delivery and effective triage of how the users arrive.

*3.1.2 Services Offered: Direct and Indirect.* When planning for a change in user experience, identifying all the services is a primary step. These services can be classified into the following categories:

- **Direct Services (User-facing)**: Tangible resources which the user interacts with immediately, such as HPC cluster interfaces, service portals, storage services (scratch/project), computational infrastructure (CPU/GPU/xPUs), application software, and direct hosted science gateways.
- **Indirect Services (RCC facing)**: The "invisible" layer running in RCC that affects user accessibility. In-house utility tools, security infrastructure and services extended through RCC collaboration with other organizations, for e.g., an Internet2 subscription may provide 'Eduroam' access to all university members. Other examples, such as Active Directory (AD) services provided by the RCC parent department or organization.
- **Latent Services (Utility)**: The latent set of services includes other utility services to empower user productivity such as: consultation, training materials and repositories, grant writing support, training workshops and seminars, hackathons, etc.

Latent and mission-critical services are often oblivious to end users, however, some of them serve as the spinal cord for the RCC itself. Hence, identifying them early on is extremely critical for planning, development and rollout phases in later stages.

*3.1.3 Quantify and Evaluate Current State.* Deciding on key KPIs and formulating a data-driven strategy has been a industry standard for decades. Nonetheless, quantifying UX at an RCC can be challenging if there is no infrastructure for it. Using technologies that support analytics, capture user clicks, session timing and journey mapping features can often capture drops and friction points in user experience. Hence, quantifying and analyzing a bigger picture can reveal patterns and help in mapping areas of poor user engagement with technological bottlenecks.

## 3.2 Data Curation

Systematic data collection, accounting, and analytics quantify an RCC's high-impact research cataclysm role. Tracking the growth of specific user groups and their software usage often plays a vital role in linking HPC usage to research output acting as a testament to a great user experience. Below, we share strategies for deriving business analytics and practices for storing them for informed decision making.

*3.2.1 Business Analytics.* The primary motivation for collecting data is to provide a bird's-eye view of RCC services and to understand patterns of RCC users' usage to further improve their experience. Such business analytics will help leadership and planning committees make informed decisions and target areas with the maximum user footprint or usage. Creating dashboards for software usage, job-level heuristics, allocation-wise active usage, and classification of batch-job errors with respect to software and applications are some examples. Collecting data on user affiliations and professional information can also reveal unexpected patterns of cross-disciplinary activity. Advanced RCCs correlate job accounting data with bibliometric databases (e.g., Web of Science, Scopus) and grant awards. This allows centers to quantify computing time in terms of publications with respect to federal and departmental funding. This "impact-based accounting" is often the deciding factor in securing institutional subsidies and renewing federal infrastructure grants [2]. Another important usage metric is 'license usage'. It allows RCCs to identify underutilized licenses, facilitating the "right-sizing" of agreements and renewals that lead to potential future cost savings.

*3.2.2 Data Management.* Many previous scholarly works provide insight into the importance and impact of collecting telemetry data at an HPC center [7, 28]. Frameworks such as "MELT" combine metrics with telemetry to identify user usage patterns [21]. However, our framework goes beyond telemetry data. We believe in the 'Data Triangulation' approach that precedes the monitoring and collecting phase and covers data beyond telemetry. We believe that data from customer-facing portals plays a critical role in providing UX analytics. Surveys often complement this step, but have a thin hit rate. Identifying mechanisms to save, click paths, and session duration are some techniques used by most cloud and social media service providers [21] that operate under strict Service Level Agreements (SLAs).

In the context of storing this analytics data, our framework adheres to a simple tenet: "Knowledge consuming more time to fetch than help in taking action is not worth the effort". Systematic, chronological storage is fundamental to extracting actionable intelligence and ensuring operational efficiency. Chronological ordering serves as the backbone for time-series analysis [25, 36]. Long-term retention enables the discovery of high-level patterns, such as seasonal anomalies or multi-year shifts in user behavior, which are often invisible in real-time operations. To support this, an archival strategy to ensure that "cold" data remains immediately accessible for compliance and retrospective study without clogging the primary data storage utility is an effective approach [31].

## 3.3 Planning and Orchestration

The user experience in any RCC is the final test of leadership and their decisions. Throughout year, the entire staff works hard to keep millions of moving parts in the center running smoothly to provide unfettered access to RCC services. Hence, in this part of the framework, we intend to provide assistance and a checklist for the RCC leadership, the core governing body that lays the path for the center's future.

*3.3.1 Objectives and Goals.* Setting organizational objectives, target goals, and timelines is not new [5]. However, in the context of improving RCC UX, we provide a guide for setting priorities and developing plans. We hypothesize tents, based on our learning and experiences that shall assist the RCC leadership to define objectives by assuming the following about users in an academic setting:

- RCC users always want more than one entry point for software and hardware resources.
- Users want a computing experience as similar as their local workstation.
- There will always be a set of users who do not like the current state affairs.
- The user will never remember training material that is unappealing to them.
- A GUI portal is never completely disliked.

Oftentimes engineering staff at RCCs rarely interact with users. Hence, we hope that the above tents can also help translate user presumptions to such teams that are less or not involved with users on regular basis.

*3.3.2 Budgeting and Discussion.* HPC centers are constantly evolving their underlying hardware and software services. Millions of dollars are spent on funding the scientific service life cycle. Formal terms and calculations, such as TCO (total cost of ownership) and ROI (return on investment) [19] are common in the RCC business model. These existing methods and studies recognize dedicated financial allocations for Human capital, software, hardware, infrastructure, energy, and other similar aspects related to the RCC itself. However, currently, there are no financial provisions or classifications dedicated to improve the user experience. The financial and non-financial costs involved in developing training and technologies to educate users are often obscured within the existing categories of the RCC business model. The cost and energy used to develop indirect services discussed in Section 3.1 are also discounted in a similar manner. Therefore, our framework seek to voice for a dedicated category within RCC's business model aiming at UX improvement.

*3.3.3 Architecture & Licensing.* Designing the UX architecture for an HPC center requires balancing various opposing forces: the immense complexity of the underlying infrastructure, security, compliance and the cognitive simplicity required for efficient use of interface. Therefore, architectural planning should also account for any necessary licenses and technology. We provide four basic principles for UXDC:

- HPC UX should eliminate the need for dependencies. The Open OnDemand portal is one great example; everything runs in a browser.
- Modern HPC UX architecture should focus on increased accessibility and adoption of science workflows rather than a single unit of compute like Jobs, GPUs, CPU, etc. The "Science Gateway" model succeeds only when it covers the "Long Tail" of science; the end-to-end process, not just the calculation step [20].
- The new UX architecture should integrate CLI-GUI integration via APIs. The Web Portal and the CLI should both call the same underlying API. This allows a user to "Start a job in CLI, Monitor it on Web, and Visualize it on Tablet."
- HPC interfaces and entry points often obscure operational complexity to facilitate accessibility. However, we propose that should be made a user's choice, and that the UX architecture should be designed with 'Operational Transparency' in mind, aligning with Nielsen's first heuristic ("Visibility of system status") which is often neglected in HPC architecture [24].

*3.3.4 Implementation and Phase-out Plan.* In HPC centers, the implementation and phase out of services are not merely technical milestones; they are critical "User Experience (UX) Events." From a UX perspective, these phases represent the highest friction points for researchers. Poor implementation of UX tools can create a "learning cliff" that discourages adoption. Similarly, a poorly planned phase-out of old services may create "migration anxiety" or data loss. Hence, it is vital to balance and time both events: the implementation of new UX and the phase out of old UX. It is important for the plan to have two faces, one for the users and the other for internal tasks. Also, note that the implementation and roll out strategies for each of the development tasks, archival and restoration of analytics data should be documented at this stage and communicated to respective stakeholders.

### 3.4 Development Phase

*3.4.1 Policies and Process.* The very first step in UXDC is developing policies and processes. Most RCCs are part of a central division in the University, such as the 'Department of Information Technology'. Hence, they are subject to the central university and departmental policies. This can be both challenging and championing for improving the UX. The most critical policy area is security. Firewalls, load balancers, and public-facing infrastructure may be subject to additional security measures and university or higher departmental policies. RCCs that have a user base working with grants from sensitive organizations such as DoD (Department of Defense), NIH, and the like must account for additional policies and security measures in place; HIPAA [34] and NIH's latest guidelines [23] are some examples.

*3.4.2 Websites and Portals.* GUIs are seldom hated by RCC users from a non-computer science background. Open-OnDemand's global acceptability and success are one such testimony. Hence, RCCs should mold and develop open-source software according to their needs. Filling blank space with something is not an ideology that this framework supports. We propose that RCCs should facilitate a UI portal based on usage patterns identified, allocated budgets, and timelines to go live. RCCs can collaborate and identify existing science gateways that work for their center. Regional computing consortia are a great place to meet local RCC staff and users to discuss patterns in region-specific usage and UX solutions.

*3.4.3 Trials, Betas and Proof of Concepts (PoCs).* Introducing software and usability change may face cultural resistance from a predominant set of practices. Hence, all development should be tested with a special set of users identified in the implementation plan (Sec. 3.3) [11]. A Beta phase allows you to recruit influential PIs (Principal Investigators), leadership personnel, and faculty who can validate tools and act as 'proof' in the new UX concept. Developing 'Beta Recruitment Emails' alongside the 'User Agreement,' which may clearly set expectations for release and usage is the preferred approach. However, PoCs and trials should be complemented with a mechanism for observability and feedback to quantify impact of the change [3, 17]. Additionally, adhering to policies or taking exceptions is a step often missed or skipped under the excuse of 'beta testing'.

*3.4.4 Training Resources.* Many previous attempts have been made for 'perfect training resources'. Micro-credentials [6] for users, interactive courses, video repositories, and an innovative learning platform for documentation are some recent examples [12]. However, there is no panacea or a perfect set of training resources due to the high cardinality of the worldwide HPC user base. Centers should constantly find new ways to train their users. However, we advocate the tents described in Sections 3.3 and 3.1. All training resources must not depend on the software environment, and the training experience should closely resonate with the user's local workstation.

### 3.5 Deployments and Release

Our framework categorizes steps involved in deployment and release of services into distinct category. This process is initiated through comprehensive communication strategies that ensure the diverse RCC user set is informed of upcoming changes and maintenance windows.

*3.5.1 Communications.* RCC should engage in pre-deployment communication and stakeholder engagement to prevent a gap between systems administration and end-user productivity. Communication roll out strategy should be developed like a story of changes from a user's perspective. The chronological sequence of events and their impact on the user is what matters. A well-thought-out plan and proposed (user-facing) architecture should be transparent to users and notified in accordance with timelines. Transparency is essential for maintaining the "Availability, Accessibility, and Usability" (AAU) metrics that define modern supercomputing democratization [30]. Communications should be crafted for each tool or platform leveraged in RCC. With the advent of modern generative AI tools, we believe it is now time to retire text-based communication in favor of more graphic-based material. Lastly, regular and timely reminders of critical deadlines should also be pre-planned in communication strategy.

*3.5.2 Service Upgrades.* Upgrading HPC services is no longer a hardware endeavor. With a rich set of software stacks and ever-evolving complex networking and storage solutions, HPC admins have to take a 'swiss army knife' approach to service upgrades. Timing, versions, and ordering in upgrades matter at centers working with thousands of computational nodes, software modules, and integrations with Single Sign-On (SSO) for authentication. The challenge becomes even steeper when compliance, vendor, and additional governmental policies are considered, especially at national labs and similar venues. Hence, administrations should draw up a 'release framework' that minimizes user-level disruptions and infrequent maintenance windows.

*3.5.3 Go-Live and Releases.* The final step in deployment is to champion the timelines and prioritize organizational goals. 'Go-Live' activities must also account for software provenance. Using external platforms for documentation and client-side binaries provides a vital fail-safe, ensuring researchers can access critical support resources even during primary system maintenance windows.

## 3.6 Feedback and Support

Retrospective analysis in HPC often focuses on identifying "cognitive friction" points where the system forces the user to divert attention from their research to handle administrative or technical overhead. Releasing products and improvements to users does not marks the end of UX improvement journey. In the context of cyberinfrastructure, *User-Centeric Design* provides the foundational theoretical framework for this retrospection. Studies have shown that even experienced researchers can be demotivated by the steep learning curve of parallel programming and job management [32]. Hence, RCC leadership should also focus on strategies to capture feedback, monitor user engagement and develop strategies to engage with PIs and researchers.

*3.6.1 Feedback.* In the HPC context, collecting UX feedback that is in alignment with previously established data analytics and storage strategies often providing meaningful insights. Training sessions and specialized courses that provide micro-credentials is a good strategy to increase user engagement [6]. These credentials provide a way for researchers to demonstrate proficiency in specialized domains such as GPU programming, parallel computing with MPI, or ethics in AI. Moreover, engaging with 'User Advisory Groups' for feature prioritization is often a most practical and direct feedback.

*3.6.2 Support.* There is no question to the fact that continued support is a critical piece in user experience. Office hours or portal requests are structured consulting mechanisms. We argue that in addition to IaaS, SaaS; RCCs also provides "Consulting-as-a-service" (CaaS) which *should be* put in the same tier as other service offerings since CaaS consume a chunk from the same pile of resources, finance and operations. Furthermore, by implementing post-release pipelines, communications and establishing timed-windows for patching, HPC centers can significantly reduce the cognitive friction associated with CI/CD cycles in scientific computing [13, 29].

## 4 User Onboarding Case Study At RIS

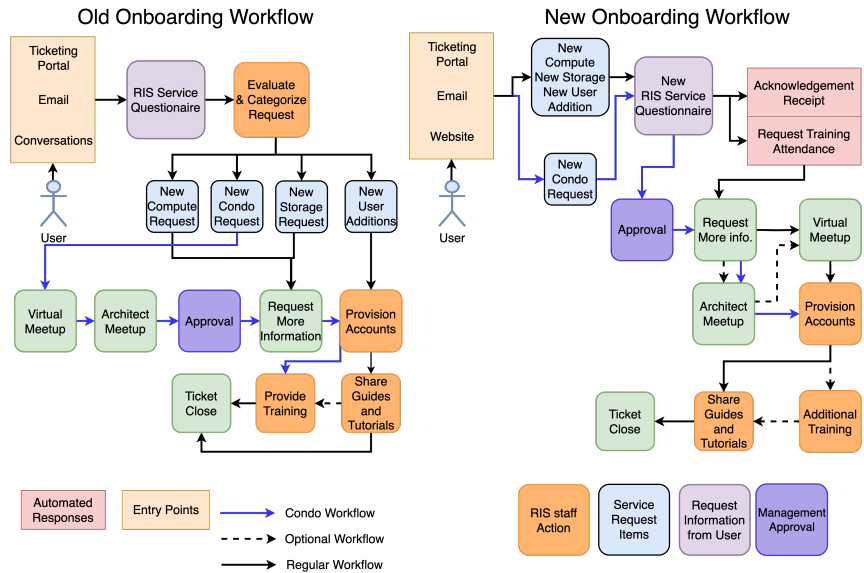

Fig. 2. Old Vs. New Onboarding Workflow

In Fall 2025, we launched our new Compute2 cluster with a completely new onboarding experience with entire 'Onboarding Revamp' project ran for entire year with a 'User-Centric' approach. We analyzed and quantified the current challenges users experience through extensive data analytics and strategies. Finally, we formulated an implementation plan to roll out changes in our internal processes, policies, and various software infrastructure used by RIS users.

### 4.1 Policy Changes

With our new Compute2 cluster launch in Fall 2025, we rolled out several policy changes intended to accelerate changes in our operating process aimed toward better user experience. Several new policies such as VPN-free access to 'Open-OnDemand' portal, increased SSH idle-time outs, mandatory training requirement for account access, setting up of minimal viable product were laid out. These policies had direct and indirect affect on user experience as they detemine the course of technical architecture. Table 4 shows impact of these changes across our services and to the end user.

### 4.2 Website & Ticketing Portal Integration

To improve the onboarding experience, our website and ticket portal were redesigned to deliver a more integrated, user-friendly interface across all entry points, as noted in Section 3.1. This effort involved a comprehensive technical overhaul to help users more easily start and continue their journey with RIS. Portal interfaces and website entry points were integrated with JIRA APIs for a strong focus on accessibility, speed, and ease of use. Additionally, our goal was to reduce friction and make key information easier to find. For that, we completely changed our ticket portal from having a 'service-oriented' design to 'user-oriented'. Figure 3 shows the contrast for service-request categorization in the old and new portal. These initiatives led to reduce response time and incorrect labeling of user requests.

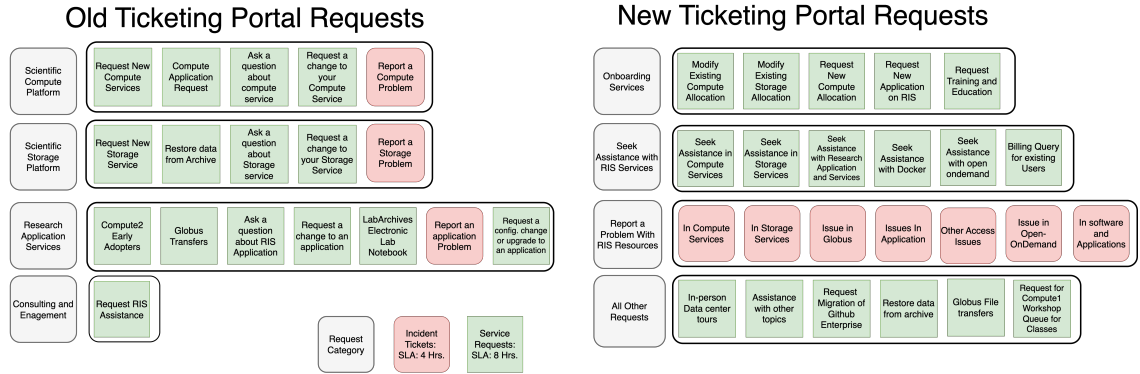

Fig. 3. User Request categorization in Service-oriented old portal Vs User-oriented new portal.

### 4.3 New Documentation Portal

Previously, RIS user-facing documentation was a standalone website lacking analytics capabilities and demanded development infrastructure and efforts for regular updates. Consequently, the documentation was migrated to 'Confluence' removing development hurdles, streamlining access management and eliminating local configuration dependencies while promising feature integrity. Confluence also offered superior integration with other atlassian products used at RIS, specifically with the JIRA service desk. This integration enabled 'ROVO AI assistant' which is capable of answering from knowledge base articles in Confluence and raising service desk requests in JIRA that. Table 1 show metrics for AI assistance and documentation portal before and after launch of Compute2 cluster. As evident, with new cluster and its documentation, there is drop in old cluster's view of compute Ultimately, this integration supports a more holistic onboarding process and provides an even more extensive experience to RIS services for new users.

Table 1. Rovo AI and Documentation Metrics Before and After Compute2 Launch

| Component | Item | Before (Compute2 Launch) | After |
| --- | --- | --- | --- |
| Rovo AI | AI-assisted Queries Resolution | 28 | 10 |
| | AI-assisted Queries Escalation | 70 | 67 |
| | AI-assisted Queries Non-Resolution | 84 | 116 |
| Documentation | Compute Cluster 1 | 1106 | 816 |
| | Compute Cluster 2 | N/A | 1454 |

### 4.4 Creation of Mission Critical Team

With a goal to enhance user support and reduce response times and friction for user queries, RIS leadership created a dedicated 'Mission Critical Team' that included members from across teams with different skills and expertise. The primary responsibilities were handling outages, complex user tickets and reduce friction in ticket resolution that demanded triage from different teams in the department.

## 5 Results & Conclusion

Improving the user experience is a cycle. Table 2 shows the impact of our changes made to website, portal and policies respectively for the last year. We measured KPIs for a period of three months each, before and after rolling out these changes with our 'Compute2' cluster. As evident, we see an increased footprint at our website. Similarly, improvements in office availability by RIS staff and website improvements resulted in increased engagement in office hours. With ticketing portal improvements, we see a drop in incorrect request categorization in general and for onboarding requests. Customer satisfaction also increased due to automated follow ups and standardized responses which also played a part in reduction in SLA times for 'First response and Resolution'. Departmental policy changes and creation of mission critical team are other additional factors that contributed to reduction in these times. Moreover, the leadership was also pleased to see improved communication skills and camaraderie across department.

Table 2. Impact of Various Changes Done in Website and Ticketing Portal During User Onboarding Experience

| Component | KPI | Before (1 Apr–1 Jul) | After (1 Sep–1 Dec) |
| --- | --- | --- | --- |
| Website | Onboarding Page Visits – Average Monthly | N/A | 941 |
| | Office Hours Page Visits – Average Monthly | 213 | 317 |
| | Booked Office Hours Sessions | 75 | 73 |
| | Cost Calculator – monthly downloads | N/A | 30 |
| | User Sessions – monthly average | 1561 | 1934 |
| Ticketing Portal | Incorrect Request Categorization | 29/115 (25%) | 30/224 (14%) |
| | Incorrect Onboarding Requests | 11 | 14 |
| | Avg. customer satisfaction | 4.9/5 | 4.96/5 |
| | Number of outages | N/A | N/A |
| Policy | SLA Avg. Time to First Response (Incident/SR) | 5.93 Hrs / 7.02 Hrs | 4.4 Hrs / 2.38 Hrs |
| | SLA Avg. Time to Resolution (Incident/SR) | 1.92 Days / 2.34 Days | 1.49 Days / 22.49 Hrs |

Through this work we aim to share a robust framework designed to serve as a guiding lamp for those determined to enhance user experience at their HPC centers. We hope that our experience and insights gained through our journey shall provide a valuable resource to the wider research community.

## A Appendix A

### A.1 Use of Generative AI

The work uses generative AI platform Gemini-AI provided to WashU in partnership with Google for correcting grammar, syntax and spelling errors. Gemini was also used to find a few scholarly articles and publications for the section 2 to reduce human labor intended in searching the vast internet. Gemini-AI partnered exclusively with WashU does not use or curate any data for improving the Gemini-AI.

### A.2 Tables

| School | Sponsors | Departments |
| --- | --- | --- |
| MEDICAL | 979 | 111 |
| ARTSCI | 142 | 14 |
| ENG | 104 | 8 |
| OLIN | 50 | 2 |
| BROWN | 19 | 1 |
| LAW | 2 | 2 |
| SamFox | 2 | 1 |
| Total | 1298 | 139 |

Table 3. Sponsor Counts by School

Table 4. Policy Changes, Expected User Impact, and Effected RIS Components

| Policy Changes | End User Impact | Direct/Indirect RIS Component |
| --- | --- | --- |
| Improved Access Controls | Smoother User Authentication via Duo; Reduced Session Disruption; Increased User Sessions | Command line SSH session; Open OnDemand portal; Operational Processes |
| Increased SSH Timeouts | Reduced Session Disruption; Increased User Sessions | Operational Processes; Open OnDemand portal |
| Mandatory Compute2 Training | Better User Training; Increased User engagement | Open OnDemand portal; Operational Processes |
| Multi-Instance GPUs in Interactive Queues | Increased GPU availability; Reduced Wait times; Increased Accessibility | Operational Processes; Website; Documentation Portal |
| VPN-free Portal Access | Increased User Accessibility; More User Engagement | Website; Documentation Portal; Security & Authentication |
| Mandatory Setup Training | Reduced Time to First Job; Increased Accessibility | Website; Documentation; Ticketing portal |
| Preemption Queues | Increased Resource Availability; Reduced Wait times | Command line; Operational Processes; Documentation |
| Virtual Meetups | Better User Training; Less tickets for basic queries | Office Hours; Ticketing Portal; Documentation |